\documentclass[a4paper,11pt]{article}
\usepackage[left=3.17cm,right=3.17cm,top=2.54cm,
headheight=0.5cm,headsep=0.54cm,bottom=2.54cm,footskip=0.79cm
]{geometry}

\usepackage{amsmath,amssymb,dsfont}

\usepackage{hyperref,cite}

\usepackage{graphicx}

\begin{document}
\title{Deformation quantization for systems with second-class constraints in deformed fermionic phase space}

\author{Bing-Sheng Lin$^{1,2,\dag}$,\,\,Tai-Hua Heng$^{\,3}$\\
	\small $^{1}$School of Mathematics, South China University of Technology, Guangzhou 510641, China\\
	\small$^{2}$Laboratory of Quantum Science and Engineering,\\
	\small South China University of Technology, Guangzhou 510641, China\\
	\small $^{\dag}$Email: sclbs@scut.edu.cn\\
	\small $^{3}$School of Physics and Optoelectronic Engineering, Anhui University, Hefei 230601, China
}

\date{30 May 2022}

\maketitle

\begin{abstract}
In order to quantize systems involving second-class constraints, one
should use Dirac bracket instead of Poisson bracket. Furthermore, one
can specify a star product in which the term linear in $\hbar$ is
proportional to the Dirac bracket. In this way an oscillator system in a
deformed fermionic phase space is analyzed and the corresponding energy
level and Wigner functions are evaluated according to scheme of
deformation quantization. We also study the entanglement entropy induced by the deformation of the fermionic phase space.
\\

\textit{PACS:} 03.65.-w, 02.40.Gh, 05.30.Fk
\\

\textit{Keywords:} deformed fermionic space; deformation quantization; entanglement entropy

\end{abstract}

\section{Introduction}\label{sec1}
Usually there are three logically autonomous alternative
approaches to quantum mechanics. The first one is the standard
operator formalism in Hilbert space, developed by Heisenberg,
Schr\"{o}dinger, Dirac and others in the twenties of the past
century. The second one relies on path integral, and was conceived
by Dirac and constructed by Feynman. The third one is deformation
quantization formalism \cite{s1}, based on Wigner's
quasi-distribution function \cite{s2} and Weyl's correspondence
between quantum operators and ordinary $c$-number phase space
functions \cite{s3,s31,s32,s4,s5,s6}.

Physical applications of the deformation quantization method have
mainly been restricted to systems involving bosonic degrees of
freedom, both in quantum mechanics and quantum field theory.
Some researchers also studied fermionic algebras in quantum mechanics by deformation quantization methods \cite{s7,s71,s8,s81,s82}. These researches show that
deformation quantization is a powerful tool for treating systems
involving bosonic and/or fermionic degrees of freedom.

There has been much interest in the study of physics in different kinds of deformed spaces, such as
noncommutative space \cite{s9,s10,s11,s12,s13,s14,s15}. The ideas of noncommutative or deformed space-time and field
theories defined on such a structure started already in 1947 \cite{s16}.
It came up again first in the 1980's, when Connes
formulated the mathematically rigorous framework of noncommutative
geometry \cite{s17}. In physical theories a noncommutative
space-time first appeared in string theory, namely in the
quantization of open string \cite{s18,s181,s19}. Another field of interest,
where the noncommutativity of space-time could play an important
role, is quantum gravity \cite{s20,s201}. Also in condensed matter
physics the concept of noncommutative space-time is applied, which
is relevant for the integer quantum Hall effect \cite{s21}. On the other hand, besides the
bosonic coordinate space which can be deformed as mentioned above,
the fermionic space can also be deformed to the so-called non-anticommutative space by making
the fermionic coordinates, i.e. the Grassmann variables, not
anticommuting, but satisfying a Clifford algebra. Seiberg showed
this deformation may maintain the $\mathcal{N}=1/2$ supersymmetry and arise on
branes in the presence of a graviphoton background \cite{s22}.
Some other researchers also studied some types of deformed fermionic spaces and superspaces \cite{221,222,223,224,225,226}.
So it is significant to study physical systems in a deformed fermionic space in
deformation quantization.

In this paper we use deformation quantization technique to quantize
pseudoclassical systems involving second-class constraints in deformed fermionic
space. The paper is organized as follows. Firstly, in section \ref{sec2} we
briefly summarize the procedure of quantizing such a system in the
deformed fermionic space according to the deformation quantization. Then using this
scheme we quantize a system of two fermionic oscillators in a deformed fermionic
space and obtain the energy level and corresponding Wigner functions
for the system in section \ref{sec3}. We also study the entanglement entropy of the physical systems induced by the deformation of the fermionic spaces in section \ref{sec4}. As a comparison, in section \ref{sec5} we also use operator formalism in
Hilbert space to recalculate the entanglement entropy
of this system, and get the same result. The last section is
conclusion.

\section{Pseudoclassical systems in deformed fermionic space and their quantization}\label{sec2}
We consider a pseudoclassical system with fermionic degrees of
freedom, which described by Grassmann variables $\theta_{\alpha}$.
The Grassmann parity of a quantity $f$ is denoted by
$\varepsilon(f)$, for example, $\varepsilon (\theta_\alpha)=1$ and
$\varepsilon (\theta_\alpha \theta_\beta)=0$. The Lagrangian is a
function of the Grassmann coordinates and their velocities
$L(\theta_\alpha, \dot{\theta}_\beta)$, $\alpha,\beta=1,...,n$. The
canonical conjugate momenta are
$\pi_\alpha=\overrightarrow{\partial}\!L/\partial
\dot{\theta}_\alpha$, where $\overrightarrow{\partial}/\partial
\dot{\theta}$ denotes left derivatives for Grassmann variables. Thus
we have a $2n$-dimensional odd (or Grassmann) phase space $M$, and a
point $z$ in $M$ is written as $z=(\theta_\alpha, \pi_\beta)$. The
Hamilton function is (here sum convention is understood)
\begin{equation}
H(\theta_\alpha, \pi_\beta)= \dot{\theta}_\alpha \pi_\alpha -L.
\end{equation}

The Poisson bracket $\{F,\,G \}$ of two functions $F$ and $G$ on $M$
is a $\mathbb{Z}_{2}$-graded bilinear map with the following properties:
\begin{enumerate}\label{prop}
\item[(i)\;\;] $\{F,\,G\}=-(-)^{\varepsilon (F)\,\varepsilon
(G)}\,\{G,\,F\}$;
\item[(ii)\;] $\{F,\,GH\}=\{F,\,G\}H+(-)^{\varepsilon
(F)\,\varepsilon
(G)}\,G\{F,\,H\}$;
\item[(iii)] $\{\{F,G\},H\}+(-)^{\varepsilon(F)(\varepsilon(G)+\varepsilon(H))}\{\{G,H\},F\}
+(-)^{\varepsilon(H)(\varepsilon(F)+\varepsilon(G))}\{\{H,F\},G\}=0$.
\end{enumerate}
In the ordinary odd phase space $M$ the Poisson bracket may be
written in terms of canonical coordinates as
\begin{equation}\label{pbfg}
\{F,\,G\}= -(-)^{\varepsilon (F)}\left(
\frac{\overrightarrow{\partial}\!F}{\partial
\theta_\alpha}\,\frac{\overrightarrow{\partial}\!G}{\partial
\pi_\alpha}+\frac{\overrightarrow{\partial}\!F}{\partial
\pi_\alpha}\,\frac{\overrightarrow{\partial}\!G}{\partial
\theta_\alpha } \right),
\end{equation}
and we have the canonical relations
\begin{equation}\label{thepi}
\{\theta_\alpha,\,\pi_\beta \}=\delta_{\alpha \beta},
\end{equation}
the brackets involving two coordinates or two momentum variables
vanish. The Hamilton equations governing evolution of the
corresponding pseudoclassical system may be written as
\begin{equation}
\frac{{\rm d} F}{{\rm d} t}=\{F,\,H\}.
\end{equation}
We may abbreviate our notation in Eq.\,(\ref{pbfg}) in terms of the coordinate
$z=(z_1,...,z_{2n})$ in the $2n$-dimensional odd phase space $M$ and
a Poisson tensor $A_{ij}$, where the indices $i,j$ run from $1$ to
$2n$. In canonical variables, from Eq.\,(\ref{pbfg}) the $A$ can be read as $A=
\left(
\begin{array}{cc} 0 & I_{n} \\ I_{n} & 0 \end{array} \right)$, where
$I_{n}$ is the $n \times n$ unit matrix. Then Eq.~(\ref{pbfg}) becomes
\begin{equation}
\{F,\,G\}=- A_{ij}\,F\, \overleftarrow{\partial}_{\!z_i}\,
\overrightarrow{\partial}_{\!z_j}\,G,
\end{equation}
where the relation between the left and right derivatives of
functions of Grassmann variables has been used
\begin{equation}
\frac{\overrightarrow{\partial}}{\partial \theta_{\alpha}}\,F =
(-)^{\varepsilon (F)}\,F \frac{\overleftarrow{\partial}}{\partial
\theta_{\alpha}}.
\end{equation}
In these notations we have $\frac{\overrightarrow{\partial}}{\partial
\theta_{\alpha}}\,\theta_{\beta}=\delta_{\alpha \beta}=-
\theta_{\beta} \frac{\overleftarrow{\partial}}{\partial
\theta_{\alpha}}$.

Noticing that the above simple expression of $A$ only works for
canonical variables and in an ordinary anticommutative odd
phase space. In more complicated case, from the property (i) of the
Poisson brackets we realize that the Poisson tensor matrix $A$ for
the odd phase space must be symmetric. Therefore we may use a
suitable symmetric matrix $\mathcal{A}$ to reflect
anticommutativity of a deformed fermionic phase space, and define the corresponding
Poisson bracket of two functions $F$ and $G$ on such a deformed fermionic
phase space as
\begin{equation}
\{F,\,G\}_{\rm df}= - \mathcal{A}_{ij}\,F\,
\overleftarrow{\partial}_{\!z_i}\, \overrightarrow{\partial}_{\!z_j}\,G.
\end{equation}
Of course, such defined Poisson brackets must satisfy the above
properties (i), (ii) and
(iii). According to this definition, the
canonical relations between $\theta_\alpha, \pi_\beta$ should be
different from Eq.~(\ref{thepi}) in general to reflect the
commutativity of the deformed fermionic phase space, and the Hamilton
equation of motion for a pseudoclassical system on such a deformed fermionic
phase space is still
\begin{equation}
\frac{{\rm d} F}{{\rm d} t}=\{F,\,H\}_{\rm df}.
\end{equation}

For a classical system involving constraints additional concepts
arise. Usually the constraints can be divided into two classes. The
first-class constraints are those whose Poisson brackets with all
other constraints vanish, and the remaining constraints are
second-class. According to Dirac \cite{s23}, to describe systems
with second-class constraints $\chi_{\alpha}$, $\alpha=1,...,l$ in
the deformed fermionic phase space, one should use Dirac brackets instead of the
Poisson brackets,
\begin{equation}\label{dbfg}
\{F,\,G\}_{\rm D}=\{F,\,G\}_{\rm df}-\{F,\,\chi_{\alpha}\}_{\rm df}\,C^{-1}_{\alpha
\beta}\{\chi_{\beta}, \,G\}_{\rm df},
\end{equation}
where $C_{\alpha \beta}=\{\chi_{\alpha},\,\chi_{\beta}\}_{\rm df}$ and
$C^{-1}_{\alpha \beta}$ is the matrix inverse to $C_{\alpha \beta}$.
Then the second-class constraints should be set strongly to zero.

In order to quantize the pseudoclassical systems living in such a
deformed fermionic phase space, by the procedure of deformation quantization, the
ordinary multiplication of the pseudoclassical functions on the deformed fermionic
phase space should be replaced by a deformed associative
$*$-product. The $*$-product should be chosen in such a way so that
the term linear in the deformation parameter (for example, $\hbar$)
in the $*$-product is proportional to the Poisson bracket. In the
case of systems involving second-class constraints the term linear
in $\hbar$ in the $*$-product should be proportional to the Dirac
bracket instead of the Poisson bracket. The $*$-product on a given
phase space is in the best case unique only up to an equivalence
class. Equivalent $*$-products correspond to different quantization
schemes. Usually one use the Moyal $*$-product, whose general form
in the deformed fermionic phase space (when the second-class constraints do not
involve) should be
\begin{equation}
F*G=F\,\exp \left( -\frac{{\rm i}\hbar}{2}
\mathcal{A}_{ij}\,\overleftarrow{\partial}_{\!z_i}\,
\overrightarrow{\partial}_{\!z_j}  \right)\,G.
\end{equation}

The so-called time-evolution function $\mathrm{Exp}(-{\rm i}Ht/\hbar)$ is the solution of the equation
\begin{equation}
{\rm i}\hbar \frac{{\rm d}}{{\rm d} t} \mathrm{Exp}(-{\rm i}Ht/\hbar) =
H*\mathrm{Exp}(-{\rm i}Ht/\hbar),
\end{equation}
which is the equivalent of the time-dependent Sch\"{o}dinger
equation in this context. For a time-independent Hamilton function
the solution is given  by the star exponential
\begin{equation}
\mathrm{Exp}(-{\rm i}Ht/\hbar)=\sum_{n=0}^{\infty} \frac{1}{n!} \left(
\frac{-{\rm i}t}{\hbar} \right)^{n}\, H_{*}^{n},
\end{equation}
where $H_{*}^{n}$ is the $n$-th $*-$power of the function $H$,
\begin{equation}
H^n_{*}=\underbrace{H* H *...* H}_n.
\end{equation}
The
time-evolution function has a Fourier-Dirichlet expansion of the
form\cite{s6}
\begin{equation}
\mathrm{Exp}(-{\rm i}Ht/\hbar)=\sum_{E} W_{E}\, {\rm e}^{-{\rm i}Et/\hbar},
\end{equation}
the $W_{E}$ are Wigner functions corresponding to the energy $E$,
which satisfy the $*$-genvalue equation
\begin{equation}
H*W_{E}=E\,W_{E}=W_{E}*H.
\end{equation}
This $*$-genvalue equation is the equivalent of the
time-independent Schr\"{o}dinger equation in this context. The
spectral decomposition of the Hamilton function is given by
\begin{equation}
H=\sum_{E} E\,W_{E}.
\end{equation}
The Wigner functions are idempotent and complete
\begin{equation}
W_{E}*W_{E'}=\delta_{E,E'}\,W_{E},~~~~~~~~\sum_{E}W_{E}=1.
\end{equation}
In the following sections we show the concrete procedures of
deformation quantization by discussing a simple example of two
fermionic oscillators in a special type of deformed fermionic space.

\section{Two fermionic oscillators in a deformed fermionic space}\label{sec3}
Some authors are
interested in a case that the anticommuting variables form a
Clifford algebra like \cite{s22,221,222,223,224,225,226}
\begin{equation}\label{cliff}
\{ \theta_{\alpha},\,\theta_{\beta} \}_{*}=c_{\alpha \beta},
\end{equation}
where $\{F,\,G \}_{*}\equiv F*G+G*F$ is a star anticommutator, and this
deformed algebra may arise in string theory in a graviphoton background.
We can call this space non-anticommutative (NAC) space.

Consider a fermionic system in the NAC space involves four
Grassmann coordinates $\theta_\alpha$, $\alpha=1,2,3,4$. The
Lagrange function is
\begin{equation}
L=\frac{{\rm i}}{2}( \theta_{1} \dot{\theta}_{1} + \theta_{2}
\dot{\theta}_{2} + \theta_{3} \dot{\theta}_{3} + \theta_{4}
\dot{\theta}_{4} ) + {\rm i}\omega \theta_{1}\theta_{3} +
{\rm i}\omega \theta_{2}\theta_{4},
\end{equation}
which describes a system of two fermionic oscillators in the NAC
space. The canonical momenta are $\pi_\alpha=-\frac{{\rm i}}{2}
\theta_\alpha$, so that the constraints are $\chi_\alpha=
\pi_\alpha+ \frac{{\rm i}}{2}\theta_\alpha$, and the Hamilton
function is given by
\begin{equation}\label{ham}
H=\dot{\theta}_\alpha \pi_\alpha -L=-{\rm i}\omega
\theta_{1}\theta_{3} -{\rm i}\omega \theta_{2}\theta_{4}.
\end{equation}
For example, one can use the following Poisson bracket in the non-anticommutative space,
\begin{eqnarray}
\{ F,\,G \}_{\rm nac} &=&-F\Big[\overleftarrow{\partial}_{\theta_\alpha}
\overrightarrow{\partial}_{\pi_\alpha} +
\overleftarrow{\partial}_{\pi_\alpha}
\overrightarrow{\partial}_{\theta_\alpha} \nonumber\\
&&~+ {\rm i} C \left(
\overleftarrow{\partial}_{\theta_{1}}
\overrightarrow{\partial}_{\theta_{2}} +
\overleftarrow{\partial}_{\theta_{2}}
\overrightarrow{\partial}_{\theta_{1}} +
\overleftarrow{\partial}_{\theta_{3}}
\overrightarrow{\partial}_{\theta_{4}} +
\overleftarrow{\partial}_{\theta_{4}}
\overrightarrow{\partial}_{\theta_{3}} \right) \Big] G,
\end{eqnarray}
where $C$ is some parameter.
It is easy to verify that, this Poisson bracket satisfies the properties (i),
(ii) and (iii)
in the previous section. Thus the Poisson brackets of the
constraints are
\begin{equation}
\{ \chi_\alpha,\,\chi_\beta \}_{\rm nac}={\rm i}\delta_{\alpha \beta}
- {\rm i}\frac{C}{4} \left( \delta_{\alpha 1}\delta_{\beta 2}
+\delta_{\alpha 2}\delta_{\beta 1}+\delta_{\alpha 3}\delta_{\beta
4} +\delta_{\alpha 4}\delta_{\beta 3}\right),
\end{equation}
which mean that the constraints are second-class, and they can be
expressed by a matrix
\begin{equation}
C_{\alpha \beta}=\{ \chi_\alpha,\,\chi_\beta \}_{\rm nac}={\rm i}
\left(
\begin{array}{cccc} 1 & -\frac{C}{4} & 0 & 0 \\ -\frac{C}{4} & 1 & 0 &
0 \\ 0 & 0 & 1 & -\frac{C}{4} \\ 0 & 0 & -\frac{C}{4} & 1
\end{array} \right)
\end{equation}
with its inverse
\begin{equation}
C^{-1}_{\alpha \beta}=\frac{-{\rm i}}{1-\frac{C^2}{16}} \left(
\begin{array}{cccc} 1 & \frac{C}{4} & 0 & 0 \\ \frac{C}{4} & 1 & 0 &
0 \\ 0 & 0 & 1 & \frac{C}{4} \\ 0 & 0& \frac{C}{4} & 1
\end{array} \right).
\end{equation}
Using Eq.~(\ref{dbfg}) we obtain the Dirac brackets in this context,
\begin{eqnarray}
\{  F,\,G \}_{\rm D}&=&  \frac{4{\rm i}}{1-\frac{C^2}{16}} F \left(
\overleftarrow{\partial}_{\theta_{1}}
\overrightarrow{\partial}_{\theta_{1}} +
\overleftarrow{\partial}_{\theta_{2}}
\overrightarrow{\partial}_{\theta_{2}} +
\overleftarrow{\partial}_{\theta_{3}}
\overrightarrow{\partial}_{\theta_{3}} +
\overleftarrow{\partial}_{\theta_{4}}
\overrightarrow{\partial}_{\theta_{4}} \right) G \nonumber\\
& &~ + \frac{{\rm i}C}{1-\frac{C^2}{16}}  F \left(
\overleftarrow{\partial}_{\theta_{1}}
\overrightarrow{\partial}_{\theta_{2}} +
\overleftarrow{\partial}_{\theta_{2}}
\overrightarrow{\partial}_{\theta_{1}} +
\overleftarrow{\partial}_{\theta_{3}}
\overrightarrow{\partial}_{\theta_{4}} +
\overleftarrow{\partial}_{\theta_{4}}
\overrightarrow{\partial}_{\theta_{3}} \right) G.
\end{eqnarray}
Here we have implemented the constraints as strong equations
according to Dirac's method\cite{s23} so the only independent
variables are then the $\theta_{\alpha}$. From now on we use the
Dirac brackets instead of the Poisson brackets in this NAC space.

As mentioned in section \ref{sec2}, in order to quantize a classical system the
Moyal star product should be chosen so that its first-order term is
proportional to the Dirac bracket in the case involving second class
constraints. Thus we use the following Moyal star product for the NAC
space
\begin{equation}\label{starnac}
*'=\exp \left( \frac{1}{2} \overleftarrow{\partial}_{\theta_{\alpha}}
\mathcal{A}_{\alpha \beta}
\overrightarrow{\partial}_{\theta_{\beta}} \right),
\qquad \alpha,\beta=1,2,3,4,
\end{equation}
where the symmetric matrix $\mathcal{A}$ is
\begin{equation}
\mathcal{A}=  \left( \begin{array}{cccc}
\hbar & c & 0 & 0 \\
c & \hbar & 0 & 0 \\
0 & 0 & \hbar & c \\
0 & 0 & c & \hbar
\end{array} \right),
\end{equation}
with $c=\hbar \,C /4$.
It is easy to verify that the independent
variables satisfy the following star anticommutators,
\begin{equation}
\{ \theta_{i},\,\theta_{i} \}_{*'}= \hbar\,,\qquad \{\theta_{1},\,\theta_{2} \}_{*'}=\{ \theta_{3},\,\theta_{4}\}_{*'}=c,\qquad i=1,2,3,4,
\end{equation}
and others vanish.

Furthermore, in the present work, we will consider a more general case, in which the independent
variables satisfying the following star anticommutators
\begin{equation}
\{ \theta_{i},\,\theta_{i} \}_{*}=\hbar\,,\qquad
\{\theta_{1},\,\theta_{2} \}_{*}=c\,,
\quad
\{ \theta_{3},\,\theta_{4}\}_{*}=d,\qquad i=1,2,3,4,
\end{equation}
and others vanish.
The parameters $c, d$ are real numbers, and usually we also assume $|c|, |d| \ll \hbar$. The corresponding star product is
\begin{eqnarray}
F*G &=&F\,\exp\!\left[\frac{\hbar}{2} \left(
\overleftarrow{\partial}_{\!\theta_{1}}
\overrightarrow{\partial}_{\!\theta_{1}} +
\overleftarrow{\partial}_{\!\theta_{2}}
\overrightarrow{\partial}_{\!\theta_{2}} +
\overleftarrow{\partial}_{\!\theta_{3}}
\overrightarrow{\partial}_{\!\theta_{3}} +
\overleftarrow{\partial}_{\!\theta_{4}}
\overrightarrow{\partial}_{\!\theta_{4}} \right)\right.\nonumber\\
& &~~\left. + \frac{c}{2}\left(
\overleftarrow{\partial}_{\!\theta_{1}}
\overrightarrow{\partial}_{\!\theta_{2}} +
\overleftarrow{\partial}_{\!\theta_{2}}
\overrightarrow{\partial}_{\!\theta_{1}}\right) +
\frac{d}{2}\left(\overleftarrow{\partial}_{\!\theta_{3}}
\overrightarrow{\partial}_{\!\theta_{4}} +
\overleftarrow{\partial}_{\!\theta_{4}}
\overrightarrow{\partial}_{\!\theta_{3}} \right)\right] G.
\end{eqnarray}
This means that the fermionic star product leads to a
Cliffordization of the Grassmann algebra of the fermionic variables,
and it is a special realization of Eq.~(\ref{cliff}). In this stage one finds
out that if the deformation parameters $c=d=0$, the Hamilton function
(\ref{ham}) describes two independent fermionic oscillators.

For convenience, one can divide the Hamilton
function (\ref{ham}) into two parts as follow,
\begin{equation}
H=H_++H_-,
\end{equation}
where
\begin{eqnarray}
&&H_+=-\frac{{\rm i}\omega}{2}(\theta_{1}\theta_{3}+\theta_{2}\theta_{4}
+\theta_{1}\theta_{4}+\theta_{2}\theta_{3})\nonumber\\
&&H_-=-\frac{{\rm i}\omega}{2}(\theta_{1}\theta_{3}+\theta_{2}\theta_{4}
-\theta_{1}\theta_{4}-\theta_{2}\theta_{3}).
\end{eqnarray}
It is easy to verify that
\begin{equation}
H_+*H_-=H_+ H_-=H_-H_+=H_-*H_+,
\end{equation}
and
\begin{equation}
H_\pm*H_\pm=\frac{h_\pm^2\omega^2}{4}.
\end{equation}
where $h_\pm=\sqrt{(\hbar\pm c)(\hbar\pm d)}$.

After some straightforward calculations, one can obtain
\begin{equation}
{\rm i}h_\pm \frac{{\rm d}}{{\rm d} t} \mathrm{Exp}(-{\rm i}H_\pm t/h_\pm)
=H_\pm*\mathrm{Exp}(-{\rm i}H_\pm t/h_\pm),
\end{equation}
where $\mathrm{Exp}(-{\rm i}H_\pm t/h_\pm)$ are the time-evolution functions,
\begin{eqnarray}
\mathrm{Exp}(-{\rm i}H_\pm t/h_\pm)&=&\sum_{n=0}^{\infty} \frac{1}{n!} \left(
\frac{-{\rm i}t}{h_\pm} \right)^{n}\, (H_\pm)_{*}^{n}\nonumber\\
&=&W_\pm^+e^{-{\rm i}\frac{\omega t}{2}}+W_\pm^-e^{{\rm i}\frac{\omega t}{2}}.
\end{eqnarray}
and
\begin{equation}
W_\pm^+=\frac{1}{2}+\frac{H_\pm}{h_\pm\omega},\qquad
W_\pm^-=\frac{1}{2}-\frac{H_\pm}{h_\pm\omega}.
\end{equation}
The corresponding $*$-eigenvalue equations are
\begin{equation}
H_\pm*W_{\pm}^i=E_\pm^i W_{\pm}^i=W_{\pm}^i*H_\pm,\qquad i=+,-,
\end{equation}
and
\begin{equation}
E_\pm^+=\frac{h_\pm\omega}{2},\qquad
E_\pm^-=-\frac{h_\pm\omega}{2},
\end{equation}
The spectral decomposition of the Hamilton function is given by
\begin{equation}
H_\pm=E_\pm^+W_\pm^++E_\pm^-W_\pm^-.
\end{equation}
The Wigner functions are idempotent and complete,
\begin{equation}
W_\pm^i*W_\pm^j=\delta_{ij}\,W_\pm^i,\qquad W_\pm^++W_\pm^-=1.
\end{equation}
We also have
\begin{equation}
W_+^i*W_-^j=W_+^iW_-^j=W_-^jW_+^i=W_-^j*W_+^i.
\end{equation}

For the total Hamiltonian (\ref{ham}), the corresponding Wigner functions are just
\begin{equation}
W_{ij}\equiv W_+^i*W_-^j,\qquad i,j=+,-,
\end{equation}
and these Wigner functions satisfy
\begin{equation}\label{wijkl}
W_{ij}*W_{kl}=\delta_{ik}\delta_{jl}W_{ij},
\end{equation}
The $*$-eigenvalue equation is
\begin{equation}
H*W_{ij}=E_{ij} W_{ij}=W_{ij}*H,
\end{equation}
where $E_{ij}=E_+^i+E_-^j$ are the energy level of the total system.

For example,
\begin{eqnarray}\label{w1}
W_{++}&=&W_+^+*W_-^+
=\frac{1}{4}+\frac{H_+}{2h_+\omega}+\frac{H_-}{2h_-\omega}
+\frac{H_+H_-}{h_+h_-\omega^2}\nonumber\\
&=&\frac{1}{4}-{\rm i}\frac{h_++h_-}{4h_+h_-}(\theta_{1}\theta_{3}+\theta_{2}\theta_{4})
+{\rm i}\frac{h_+-h_-}{4h_+h_-}(\theta_{1}\theta_{4}+\theta_{2}\theta_{3})
+\frac{\theta_{1}\theta_{2}\theta_{3}\theta_{4}}{h_+h_-}.
\end{eqnarray}

We also have \cite{hhs}
\begin{equation}\label{trw}
\mathrm{Tr}(W_{ij})=\frac{4}{\hbar^4}\int d\theta_4 d\theta_3 d\theta_2 d\theta_1 (\star W_{ij}) =1,
\end{equation}
where $\star F$ is the Hodge dual for Grassmann monimials, which maps a Grassmann monomial with grade $r$ into a Grassmann monomial with grade $d-r$, and $d$ is the number of Grassmann basis elements,
\begin{equation}
\star (\theta_{i_1} \theta_{i_2}...\theta_{i_r})
=\frac{1}{(d-r)!}\varepsilon^{i_{r+1}...i_d}_{i_1i_2...i_r}\theta_{i_{r+1}}...\theta_{i_d},
\qquad 1\leqslant r\leqslant d.
\end{equation}
The integration is given by the Berezin integral for which we have $\int d\theta_i \theta_j=\hbar\delta_{ij}$.

\section{Entanglement of the fermionic oscillators in the NAC space.}\label{sec4}
As we have already known \cite{Lin,msg,aggm}, in noncommutative bosonic spaces, the noncommutativity of the spaces can induce some types of entanglement of the physical systems.
So it is significant to study the entanglement of physical systems in deformed fermionic spaces.

In phase spaces, the quantum R\'{e}nyi entropy can be defined as \cite{zc},
\begin{equation}\label{re}
S_\alpha(W)=\frac{1}{1-\alpha}\ln\left(\mathrm{Tr}(W_*^\alpha)\right),
\end{equation}
where $\alpha\neq 1$ is some positive parameter.
In the limit for $\alpha\to 1$, the quantum R\'{e}nyi entropy is just the von Neumann entropy,
\begin{equation}
S_1(W)=-\mathrm{Tr}( W*\ln_{*}\!W),
\end{equation}
where the $*-$logarithm is
\begin{equation}
\ln_*(f):=-\sum_{n=1}^{\infty}\frac{(1-f)^n_{*}}{n}.
\end{equation}
From (\ref{wijkl}) and (\ref{trw}), one can obtain
\begin{equation}
S_\alpha(W_{ij})=0,
\end{equation}
which means that the states $W_{ij}$ are all pure states.
By virtue of the partial entanglement entropy, the entanglement measure of $W_{ij}$ can be defined as the quantum entropy of the reduced states $W_{ij}^{(1)}(\theta_1,\theta_3)$ or $W_{ij}^{(2)}(\theta_2,\theta_4)$.

As an example, let us consider the Wigner function $W_{++}$ (\ref{w1}). The corresponding Wigner functions of reduced states in the subspaces $(\theta_1,\theta_3)$ and $(\theta_2,\theta_4)$ are
\begin{eqnarray}\label{w11}
W_{++}^{(1)}(\theta_1,\theta_3) &=& \mathrm{Tr}_{\theta_2\theta_4}(W_{++})=\frac{2}{\hbar^2}\int d\theta_4 d\theta_2 (\star W_{++})\nonumber\\
&=&\frac{1}{2}-{\rm i}\frac{h_++h_-}{2h_+h_-}\theta_{1}\theta_{3},
\end{eqnarray}
where $\mathrm{Tr}_{\theta_2\theta_4}$ is the partial trace over $\theta_2, \theta_4$, and
\begin{eqnarray}\label{w12}
W_{++}^{(2)}(\theta_2,\theta_4) &=& \mathrm{Tr}_{\theta_1\theta_3}(W_{++})=\frac{2}{\hbar^2}\int d\theta_3 d\theta_1 (\star W_{++})\nonumber\\
&=&\frac{1}{2}-{\rm i}\frac{h_++h_-}{2h_+h_-}\theta_{2}\theta_{4}.
\end{eqnarray}

Since the Wigner functions are just quasi-probability distributions and they can have negative eigenvalues.
One can verify that, using the normal definition (\ref{re}), the values of quantum R\'{e}nyi entropy of the above reduced state Wigner functions (\ref{w11}) and (\ref{w12}) can be negative. In some cases one can even obtain complex number results.
So one should not use the normal definition of von Neumann entropy in this case.
Similar to Ref.~\cite{skd}, instead one can use the abstract values of the eigenvalues $p_i$ of the Wigner functions to define the quantum entropy,
\begin{equation}
\mathcal{S}(W)=-\sum_i|p_i|\ln|p_i|.
\end{equation}

Now let us define the following functions
\begin{equation}
f_1=\frac{1}{2}-\frac{{\rm i}}{\hbar}\theta_{1}\theta_{3},\qquad
f_2=\frac{1}{2}+\frac{{\rm i}}{\hbar}\theta_{1}\theta_{3}.
\end{equation}
It is easy to verify that these functions are complete and orthonormal,
\begin{equation}
\sum_{i=1,2}f_i=1,\qquad
f_i*f_j=\delta_{ij}f_i.
\end{equation}
After some straightforward calculations, one can obtain
\begin{equation}
W_{++}^{(1)}(\theta_1,\theta_3)*f_i=p_if_i,
\end{equation}
where
\begin{equation}\label{pp}
p_1=\frac{1}{2}+\hbar \frac{h_++h_-}{4h_+h_-},\qquad
p_2=\frac{1}{2}-\hbar \frac{h_++h_-}{4h_+h_-}.
\end{equation}
So $f_i$ are the eigenfunctions of $W_{++}^{(1)}(\theta_1,\theta_3)$, and $p_i$ are the corresponding eigenvalues.

It is easy to see that the above eigenvalues $p_1+p_2=1$, and $p_1\geqslant 1$, $p_2\leqslant 0$. This is because the Wigner functions are just quasi-probability distributions and they can have negative eigenvalues.

So the partial entanglement entropy $E_p$ of the state $W_{++}$ is
\begin{eqnarray}\label{es}
&&E_p(W_{++})\equiv \mathcal{S}\left(W_{++}^{(1)}(\theta_1,\theta_3)\right)
=-|p_1|\ln|p_1|-|p_2|\ln|p_2|\nonumber\\
&&=-\left|\frac{1}{2}+\hbar \frac{h_++h_-}{4h_+h_-}\right|\ln\left|\frac{1}{2}+\hbar \frac{h_++h_-}{4h_+h_-}\right|
-\left|\frac{1}{2}-\hbar \frac{h_++h_-}{4h_+h_-}\right|\ln\left|\frac{1}{2}-\hbar \frac{h_++h_-}{4h_+h_-}\right|\nonumber\\
&&=-\left(\frac{1}{2}+\hbar \frac{h_++h_-}{4h_+h_-}\right)\ln\left(\frac{1}{2}+\hbar \frac{h_++h_-}{4h_+h_-}\right)\nonumber\\
&&\qquad -\left(\hbar \frac{h_++h_-}{4h_+h_-}-\frac{1}{2}\right)\ln\left(\hbar \frac{h_++h_-}{4h_+h_-}-\frac{1}{2}\right).
\end{eqnarray}
One can also calculate the entanglement entropy of the states $W_{--}$, $W_{+-}$ and $W_{-+}$. There are
\begin{eqnarray}
&&E_p(W_{--})=E_p(W_{++})\nonumber\\
&&~=-\left(\frac{1}{2}{+}\hbar \frac{h_+{+}h_-}{4h_+h_-}\right)\ln\!\left(\frac{1}{2}{+}\hbar \frac{h_+{+}h_-}{4h_+h_-}\right)
{-}\left(\hbar \frac{h_+{+}h_-}{4h_+h_-}{-}\frac{1}{2}\right)\ln\!\left(\hbar \frac{h_+{+}h_-}{4h_+h_-}{-}\frac{1}{2}\right),\nonumber\\
&&E_p(W_{+-})=E_p(W_{-+})\\
&&~=-\left(\frac{1}{2}{+}\hbar \frac{h_+{-}h_-}{4h_+h_-}\right)\ln\!\left(\frac{1}{2}{+}\hbar \frac{h_+{-}h_-}{4h_+h_-}\right)
{-}\left(\frac{1}{2}{-}\hbar \frac{h_+{-}h_-}{4h_+h_-}\right)\ln\!\left(\frac{1}{2}{-}\hbar \frac{h_+{-}h_-}{4h_+h_-}\right).\nonumber
\end{eqnarray}
Here we always assume that $|c|, |d| \ll \hbar$.

When $c=d$, we have $h_\pm=\hbar\pm c$. There are
\begin{eqnarray}
&&E_p(W_{++})=E_p(W_{--})\nonumber\\
&&\qquad=-\frac{2\hbar^2-c^2}{2(\hbar^2- c^2)}\ln\frac{2\hbar^2-c^2}{2(\hbar^2- c^2)}
-\frac{c^2}{2(\hbar^2- c^2)}\ln\frac{c^2}{2(\hbar^2- c^2)},\nonumber\\
&&E_p(W_{+-})=E_p(W_{-+})\nonumber\\
&&\qquad=-\frac{\hbar^2+\hbar c- c^2}{2(\hbar^2- c^2)}\ln\frac{\hbar^2+\hbar c- c^2}{2(\hbar^2- c^2)}
-\frac{\hbar^2-\hbar c- c^2}{2(\hbar^2- c^2)}\ln\frac{\hbar^2-\hbar c- c^2}{2(\hbar^2- c^2)}.
\end{eqnarray}
When $c=-d$, we have $h_+=h_-=\sqrt{\hbar^2-c^2}$, and
\begin{eqnarray}
&&E_p(W_{++})=E_p(W_{--})\nonumber\\
&&\qquad=-\frac{\hbar+\sqrt{\hbar^2-c^2}}{2\sqrt{\hbar^2-c^2}}
\ln\frac{\hbar+\sqrt{\hbar^2-c^2}}{2\sqrt{\hbar^2-c^2}}
-\frac{\hbar-\sqrt{\hbar^2-c^2}}{2\sqrt{\hbar^2-c^2}}
\ln\frac{\hbar-\sqrt{\hbar^2-c^2}}{2\sqrt{\hbar^2-c^2}},\nonumber\\
&&E_p(W_{+-})=E_p(W_{-+})=\ln 2.
\end{eqnarray}
The second result above means that the states $W_{+-}$ and $W_{-+}$ are always maximally entangled states with any values $c=-d$.

For the special case $c=d=0$, there are $h_+=h_-=\hbar$. So $E_p(W_{++})=E_p(W_{--})=0$ and $E_p(W_{+-})=E_p(W_{-+})=\ln 2=0.693$.
This means that the states $W_{++}$ and $W_{--}$ have no entanglement.
This returns to the case of two independent fermionic oscillators in normal commutative fermionic space.

\begin{figure}[!ht]
\centering
\includegraphics[width=0.60\textwidth]{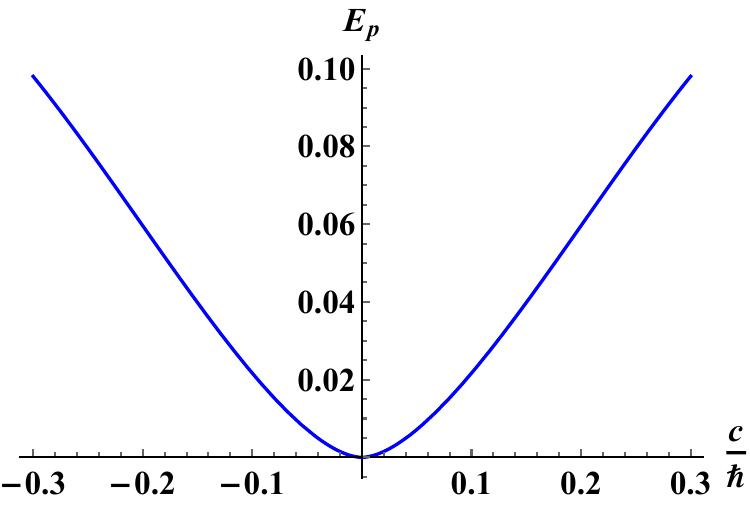}
\caption{\label{fig1}In the case $c=d$, the entanglement entropy $E_p(W_{++})=E_p(W_{--})$ with respect to the variable $c/\hbar$.}
\end{figure}

\begin{figure}[!ht]
\centering
\includegraphics[width=0.60\textwidth]{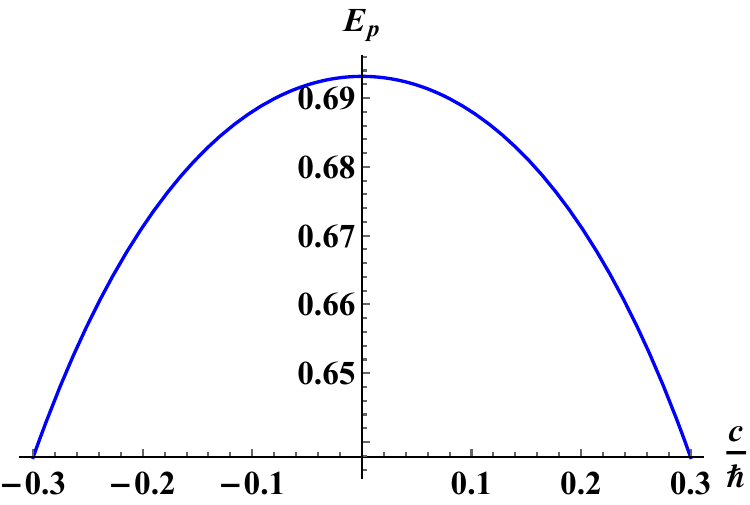}
\caption{\label{fig2}In the case $c=d$, the entanglement entropy $E_p(W_{+-})=E_p(W_{-+})$ with respect to the variable $c/\hbar$.}
\end{figure}

As an example, the entanglement entropy $E_p$ of the
states $W_{ij}$ in the case $c=d$ are plotted in Figs.~\ref{fig1} and \ref{fig2}.

Obviously, in the case $c=d$, the entanglement entropy $E_p(W_{++})=E_p(W_{--})$ becomes larger as the absolute value of $c$ increases. This means that the entanglement of the states $W_{++}$ and $W_{--}$ increases with the increase of the noncommutativity of the fermionic phase space.
On the contrary, the entanglement entropy $E_p(W_{+-})=E_p(W_{-+})$ becomes smaller as the absolute value of $c$ increases.
This means that the entanglement of the states $W_{+-}$ and $W_{-+}$ decreases with the increase of the noncommutativity of the fermionic phase space.

It is worth mentioning that, in noncommutative bosonic/fermionic phase spaces, the noncommutative coordinates can be transformed into those in normal commutative phase
space via the so-called Bopp's shift (or Seiberg-Witten maps) \cite{Gamboa}, and one can solve the new physical systems with transformed coordinates in normal commutative spaces instead of the original ones in noncommutative spaces. Similarly, one can also use such kinds of transformations in the deformed fermionic space $(\theta_1, \theta_2, \theta_3, \theta_4)$, and then study the entropy and entanglement of the two fermionic oscillator system. But usually such kinds of transformations are not unitary, which may not preserve entropies of the subsystems. So it may change the entanglement of the original system.

\section{Operator formalism in Hilbert space}\label{sec5}
As a comparison, one can also use operator formalism in
Hilbert space to analyse this system.
It is well known that bosonic Wigner operator maybe written as\cite{s2}
\begin{equation}\label{detb}
\hat{\Delta}_b (x,p)= \frac{1}{(2 \pi \hbar)^2}\int {\rm d}u{\rm d}v\,\exp
\left(\frac{{\rm i}u}{\hbar}(\hat{p}-p) +
\frac{{\rm i}v}{\hbar} (\hat{x}-x)\right),
\end{equation}
where $\hat{x}$, $\hat{p}$ are ordinary coordinate and momentum
operators (which satisfies commutation relation $[ \hat{x},\,
\hat{p} ]={\rm i} \hbar$), and $x$ and $p$ are classical canonical
coordinate and momentum variables respectively. In Eq.~(\ref{detb}) $u$ and
$v$ are real integral variables from $-\infty$ to $\infty$. Having
the above bosonic Wigner operator, for a classical function $h(x,p)$
of the canonical variables $x$ and $p$, one may get its quantum
correspondence
\begin{equation}\label{hdetb}
\hat{H} (\hat{x}, \hat{p})=\int {\rm d}p \,{\rm d}x \,h(x,p) \,\hat{\Delta}_b
(x,p).
\end{equation}
For example, if the classical function is $xp$, according to Eq.~(\ref{hdetb})
one gets its quantum correspondence $(\hat{x}\hat{p}+
\hat{p}\hat{x})/2$, which is the Weyl correspondence. Using bosonic annihilation
and creation operators $\hat{b}=(\hat{x}+
{\rm i} \hat{p})/\sqrt{2 \hbar}$ and
$\hat{b}^\dag=(\hat{x}- {\rm i}\hat{p})/\sqrt{2 \hbar}$, and holomorphic variables
$z=(x+{\rm i}p)/\sqrt{2 \hbar}$, for complex integral variable
$w=(u-{\rm i}v)/\sqrt{2}$, the above bosonic Wigner operator can
be written as
\begin{equation}
\hat{\Delta}_b (z^\ast, z)=\frac{1}{(2 \pi \hbar)^2} \int {\rm d}w^\ast
{\rm d}w\, \exp \left( \frac{w^\ast}{\sqrt{\hbar}} (\hat{b}- z) -
(\hat{b}^\dag -z^\ast) \frac{w}{\sqrt{\hbar}}\right).
\end{equation}

For a single-mode fermion, in the operator formalism the corresponding
Hilbert space is two-dimensional. The creation and
annihilation operators $\hat{a}^\dag$ and $\hat{a}$ of a single-mode
fermion satisfy the following anticommutation relations
\begin{equation}
\{ \hat{a},\,\hat{a}^\dag \}\equiv\hat{a} \hat{a}^\dag + \hat{a}^\dag
\hat{a} = 1, ~~~~~~~~\hat{a}^2=\hat{a}^{\dag 2}=0.
\end{equation}
Let $|0 \rangle$ be the vacuum
state ($\hat{a}\,|0\rangle=0$), then there is only one excitation
state $|1 \rangle = \hat{a}^\dag |0 \rangle$. The
Grassmann parities are $\varepsilon(|0 \rangle)=0$ and $\varepsilon(|1
\rangle)=1$. The operators $\hat{a}$
and $\hat{a}^\dag$ can also be expressed as
$\hat{a}=|0\rangle \langle 1 |$,
$\hat{a}^\dag=|1 \rangle\langle 0 |$.
The number
operators $\hat{a}^{\dag} \hat{a}$ acting on these states lead
to
$\hat{a}^{\dag} \hat{a} |0 \rangle = 0$,
$\hat{a}^{\dag} \hat{a} |1 \rangle = |1 \rangle$.

The pseudoclassical correspondences of the operators $\hat{a}$
and $\hat{a}^\dag$ are denoted as $\eta$ and $\eta^\ast$ respectively,
which are the Grassmann numbers. Similar to the bosonic case, the Wigner operator of a single-mode
fermion can be written as\cite{s24,s25}
\begin{eqnarray}\label{dtf}
\hat{\Delta}_f (\eta^\ast, \eta) &=& \int {\rm d}\xi^\ast {\rm d}\xi \, \exp
\left( \xi^\ast(\hat{a}- \eta)- (\hat{a}^\dag
-\eta^\ast)\xi\right)\nonumber\\
&=&\frac{1}{2} - (\hat{a}^{\dag}
-\eta^{\ast}) (\hat{a}- \eta ),
\end{eqnarray}
where $\xi$ is a complex
Grassmann integral variable. The quantum correspondence of a
pseudoclassical function $h(\eta^\ast, \eta)$ of the holomorphic
Grassmann variables $\eta^\ast$ and $\eta$ is
\begin{equation}\label{haa}
\hat{H} (\hat{a}^\dag, \hat{a}) = \int {\rm d}\eta^\ast {\rm d}\eta
\,h(\eta^\ast, \eta) \,\hat{\Delta}_f (\eta^\ast, \eta).
\end{equation}
For example, by virtue of integration laws of the Grassmann
variables, one finds that the operators  $\hat{a}$ and
$\hat{a}^\dag$ are the quantum correspondences of the holomorphic
variables $\eta$ and $\eta^\ast$, respectively.

Now let us return to discuss the system analyzed in the previous
section. One may introduce the following holomorphic variables for
our fermionic system in the NAC space
\begin{equation}\label{etathe}
\eta =\frac{1}{\sqrt{2\hbar}} \left( \theta_{1} +
{\rm i}\theta_{3}\right),\qquad
\eta^{\ast}
=\frac{1}{\sqrt{2\hbar}} \left( \theta_{1} -
{\rm i}\theta_{3}\right),
\end{equation}
which both are Grassmann variables also.
For the reduced state Wigner function $W_{++}^{(1)}(\theta_1,\theta_3)$, there is
\begin{eqnarray}
W_{++}^{(1)}(\theta_1,\theta_3) &=&\frac{1}{2}-{\rm i}\frac{h_++h_-}{2h_+h_-}\theta_{1}\theta_{3}\nonumber\\
&=&\frac{1}{2}-\hbar\frac{h_++h_-}{2h_+h_-}\eta^*\eta=W_{++}^{(1)}(\eta^*,\eta).
\end{eqnarray}
Using Eqs.~(\ref{dtf}) and (\ref{haa}), one can get its operator formalism,
\begin{equation}\label{w111}
\hat{W}_{++}^{(1)} =\int {\rm d}\eta^\ast {\rm d}\eta
\,W_{++}^{(1)}(\eta^*,\eta)\,\hat{\Delta}_f (\eta^\ast, \eta)
=\frac{1}{2}-\hbar\frac{h_++h_-}{2h_+h_-} \left( \hat{a}^{\dag} \hat{a} - \frac{1}{2} \right).
\end{equation}
Thus the two states $|0 \rangle$, $|1 \rangle$ are eigenstates of
the above operator (\ref{w111}), and the corresponding eigenvalues are
\begin{equation}
\lambda_1=\frac{1}{2}+\hbar \frac{h_++h_-}{4h_+h_-},\qquad
\lambda_2=\frac{1}{2}-\hbar \frac{h_++h_-}{4h_+h_-},
\end{equation}
which exactly corresponds to the result in Eq.~(\ref{pp}).

One can also use the matrix representations,
\begin{equation}
|0 \rangle =\left( \begin{array}{c} 1 \\ 0 \end{array}
\right),\qquad
|1 \rangle =\left( \begin{array}{c} 0 \\ 1 \end{array}
\right),
\end{equation}
and
\begin{equation}
\hat{a}=|0\rangle \langle 1 |=\left( \begin{array}{cc}
0 & 1 \\ 0 & 0 \end{array} \right),\qquad
\hat{a}^\dag=|1 \rangle
\langle 0 |=\left( \begin{array}{cc} 0 & 0 \\ 1 & 0 \end{array}
\right).
\end{equation}
So the operator (\ref{w111}) is just the following diagonal matrix,
\begin{equation}
\hat{W}_{++}^{(1)} =\left( \begin{array}{cc} \frac{1}{2}+\hbar \frac{h_++h_-}{4h_+h_-} & 0 \\ 0 & \frac{1}{2}-\hbar \frac{h_++h_-}{4h_+h_-} \end{array}
\right)=\left( \begin{array}{cc} \lambda_1 & 0 \\ 0 & \lambda_2 \end{array}
\right).
\end{equation}

It is easy to see that, for $c,d\neq 0$, there should be $\lambda_2<0<1<\lambda_1$.
So one can not use the normal formula of the von Neumann entropy. Instead, the quantum entropy of the system can be defined as $\mathcal{S}(W)=-\sum_i|\lambda_i|\ln|\lambda_i|$.

\section{Conclusion}\label{sec6}
In this paper we illustrate how to quantize systems living in a deformed fermionic space according to the procedure of deformation quantization.
Sometimes this kind of systems involves second-class constraints and
in this case the Dirac bracket is useful to replace the Poisson
bracket so that the  corresponding $*$-product can be specified. We
discuss a system of two fermionic oscillators in a so-called non-anticommutative  space,
which involves the second-class constraints. By virtue of deformation quantization methods, we obtain the Wigner functions and the corresponding energy level of
this system. We also study the entanglement induced by the deformation of the fermionic phase space. As a comparison, we also use more popular operator formalism in Hilbert space to obtain the same results, which convince ourself that the above results are correct. These efforts show
that the deformation quantization is a powerful tool to deal with
not only systems living in ordinary commutative space, but also
those living in deformed bosonic or fermionic space. We hope that these methods can help to study quantization of physical systems living in some types of deformed spaces. These results are significant to the studies of mathematical structures and physical properties of deformed spaces.

\section*{Acknowledgements}
This work is partly supported by Key Research and Development Project of Guangdong Province (Grant No.~2020B0303300001), the Guangdong Basic and Applied Basic Research Foundation (Grant No.~2019A1515011703), the Fundamental Research Funds for the Central Universities and the Natural Science Foundation of Anhui Province (Grant No.~1908085MA16).


\begin{thebibliography}{30}
\bibitem{s1} F. Bayen, M. Flato, C. Fronsdal, A. Lichnerrowicz and
D. Sternheimei, {\it Ann. Phys. (NY)} {\bf 111}, 61 (1978);
\emph{ibid.} {\it Ann. Phys. (NY)} {\bf 111}, 111 (1978).

\bibitem{s2} E. Wigner, {\it Phys. Rev.} {\bf 40}, 749 (1932).

\bibitem{s3} H. Weyl, {\it The Theory of Groups and Quantum
Mechanics} (Dover, New York, 1931).

\bibitem{s31} H. Groenewold, {\it Physica (Amsterdam)} {\bf 12}, 405 (1946).

\bibitem{s32} J. Moyal, {\it Proc. Camb. Phil. Soc.} {\bf 45}, 99 (1949).

\bibitem{s4} C. Zachos, {\it Int. J. Mod. Phys. A} {\bf17}, 297 (2002).

\bibitem{s5} T. Curtright, T. Uematsu and C. Zachos, {\it J. Math. Phys.}
{\bf 42}, 2396 (2001).

\bibitem{s6} A. Hirshfeld and P. Henselder, {\it Am. J. Phys.}
{\bf 70}, 537 (2002).

\bibitem{s7} J. C. Varilly and J. M. Gracia-Bondia, {\it Ann. Phys. (NY)}
{\bf 190}, 107 (1989).

\bibitem{s71} J. F. Carinena, J. M. Gracia-Bondia and
J. C. Varilly, {\it J. Phys. A: Math. Gen.} {\bf 23}, 901 (1990).

\bibitem{s8} A. Hirshfeld and P. Henselder, {\it Ann. Phys. (NY)}
{\bf 302}, 59 (2002).

\bibitem{s81} P.G. Castro, B. Chakraborty, Z. Kuznetsova and F. Toppan, {\it Central Eur. J. Phys.} {\bf 9} 841 (2011).

\bibitem{s82} Y. Markov and M. Markova, {\it Adv. Appl. Clifford Algebras} {\bf 31} 27 (2021).


\bibitem{s9} M. R. Douglas and N. A. Nekrasov, {\it Rev. Mod. Phys.} {\bf 73}, 977 (2001).

\bibitem{s10} M. Chaichian, P. Presnajder and A. Tureanu, {\it Phys. Rev. Lett.} {\bf 94}, 151602 (2005).

\bibitem{s11} B. S. Lin, S. C. Jing and T. H. Heng, {\it  Mod. Phys. Lett. A}
{\bf 23}, 445 (2008).

\bibitem{s12} A. Joseph, {\it Phys. Rev. D} {\bf 79}, 096004 (2009).

\bibitem{s13} B. S. Lin and T. H. Heng, {\it Chin. Phys. Lett.} {\bf 33}, 110303 (2016).

\bibitem{s14} Kh. P. Gnatenko and O. V. Shyiko, {\it Mod. Phys. Lett. A} {\bf 33} 1850091 (2018).

\bibitem{s15} N. C. Dias and J. N. Prata, {\it J. Phys. A} {\bf 52}, 225203 (2019).

\bibitem{s16} H. S. Snyder, {\it Phys. Rev.} {\bf 71}, 38 (1947);
C. N. Yang, {\it Phys. Rev.} {\bf 72}, 874 (1947).

\bibitem{s17} A. Connes, {\it Noncommutative geometry} (Academic Press, INC., 1994).

\bibitem{s18} N. Seiberg and E. Witten, {\it J. High Energy Phys.} {\bf 09}, 032 (1999).

\bibitem{s181} D. Lust, {\it J. High Energy Phys.} {\bf 12}, 084 (2010).

\bibitem{s19} C. Hull and R. J. Szabo, {\it J. High Energy Phys.} {\bf 09}, 051 (2019).

\bibitem{s20} B. M. Zupnik, {\it Class. Quantum Grav.} {\bf 24}, 15 (2007).

\bibitem{s201} M. Dimitrijevi\'{c} and V. Radovanovic, {\it Phys. Rev. D} {\bf 89}, 125021 (2014).

\bibitem{s21} A. P. Polychronakos, {\it J. High Energy Phys.} {\bf 06}, 070 (2001).

\bibitem{s22} N. Seiberg, {\it J. High Energy Phys.} {\bf 0306}, 010 (2003).

\bibitem{221} P. Castro, B. Chakraborty, Z. Kuznetsova and F. Toppan, {\it Open Phys.} {\bf 9}, 841 (2011).

\bibitem{222} A. Borowiec, J. Lukierski, M. Mozrzymas and V. N. Tolstoy, {\it J. High Energy Phys.} {\bf 06}, 154 (2012).

\bibitem{223} M. Dias, A. F. Ferrari, C. A. Palechor and C. R. Senise Jr, {\it J. Phys. A: Math. Theor.} {\bf 48}, 275403 (2015).

\bibitem{224} P. Weinreb and M. Faizal, {\it Phys. Lett. B} {\bf 748}, 102 (2015).

\bibitem{225} M. Faizal and T. S. Tsun, {\it  Found. Phys.} {\bf 45}, 1421 (2015).

\bibitem{226} C. Palechor, A. F. Ferrari and A. G. Quinto, {\it J. High Energy Phys.} {\bf 01}, 049 (2017).

\bibitem{s23} P. A. M. Dirac, {\it Lectures on quantum
mechanics} (Belfer Graduate School of Science, Yeshiva University, New York, 1964).

\bibitem{hhs} P. Henselder, A. C. Hirshfeld and T. Spernat, {\it Ann. Phys.} {\bf 317} 107 (2005).

\bibitem{Lin} B. S. Lin, J. Xu and T. H. Heng, {\it Mod. Phys. Lett. A} {\bf 34} 1950269 (2019).

\bibitem{msg} A. Muhuri, D. Sinha and S. Ghosh, {\it  Eur. Phys. J. Plus} {\bf 136} 35 (2021).

\bibitem{aggm} A. K. Armel, Y. D. Germain, T. A. Giresse and T. Martin, {\it Phys. Scr.} {\bf 96} 125731 (2021).

\bibitem{zc} C. K. Zachos, {\it J. Phys. A: Math. Theor.} {\bf 40} F407 (2007).

\bibitem{skd} P. Sadeghi, S. Khademi and A. H. Darooneh, {\it Phys. Rev. A} {\bf 86} 012119 (2012).

\bibitem{Gamboa} J. Gamboa, M. Loewe and J. C. Rojas, {\it Phys. Rev. D} {\bf 64}, 067901 (2001).

\bibitem{s24} H. Y. Fan and T. N. Ruan, {\it Commun. Theor. Phys.} {\bf 3}, 45 (1984).

\bibitem{s25} H. Y. Fan, {\it Commun. Theor. Phys.} {\bf 16}, 123 (1991).



\end{thebibliography}
\end{document}